# Construction and Evolutionary Analysis of a Game Model for Supply Chain Finance Funding Based on Blockchain Technology

Linwei Wu[1, a]*

[1] Master of Science：Project Management, Harrisburg University of Science and Technology, Harrisburg, PA 17101, USA

[a]Email: linweiwu136@163.com

*Corresponding author

**Abstract**: The current surge in supply chain finance has significantly alleviated the "capital challenges" faced by domestic related enterprises, enabling enterprises upstream and subsequent stages of the industrial chain to achieve effective circulation of financing services in the supply chain based on the credit of core enterprises. By gathering essential information from the heart of the supply chain, supply chain financing enables efficient resource distribution and aids all stakeholders in making well-informed choices. However, supply chain finance in China still faces numerous obstacles, such as information asymmetry and inefficient credit transmission chains, hindering its long-term development. This paper designs an operational framework for supply chain finance incorporating blockchain technology, clearly defines the participating entities, and analyzes their business relationships. Based upon evolutionary game theory, a supply chain finance financing game model incorporating blockchain technology is constructed. A comparative analysis of the model's equilibrium points and their stability is conducted. The choices of evolutionary equilibrium strategies adopted by small and medium-sized enterprises, key players, and financing entities within this framework are explored, and the influence of blockchain technology on the prerequisites for completing supply chain finance transactions is investigated.

## 1. Introduction

In recent times, diverse types of Chinese companies have shown robust growth, wherein private enterprises, notably small and medium-sized enterprises (SMEs), have emerged as the leading players in China's foreign trade arena and hold a crucial position in fostering economic and societal advancement. However, SMEs often face "financing challenges." The rise of supply chain finance has broken this bottleneck for them, enabling other enterprises upstream and downstream segments of the industrial chain to leverage the credit of core enterprises to achieve effective transmission of financing services within the supply chain. Through the deep integration of finance and the real economy, it has reduced operating costs across the entire chain, facilitated the construction of an industrial ecosystem characterized by mutual benefit, win-win results, and harmonious coexistence, and ensured the sustainable development of industries. As technology rapidly advances and novel innovations come to the forefront, supply chain finance has been injected with fresh energy and creative approaches. Blockchain technology, with its features of distributed ledger, tamper-resistant data, and multi-party consensus, is gradually becoming a core force driving profound changes in supply chain finance. Specifically, the application of blockchain technology makes data storage no longer reliant on a single central node but distributed across multiple nodes, achieving true decentralization, enhancing data security, reducing the risk of single-point failure, ensuring the reliability of data storage and transmission on the chain, and effectively preventing data tampering and loss. Additionally, the multi-party consensus mechanism of blockchain enables instant data sharing, breaks down information barriers, strengthens mutual trust among parties, improves transaction efficiency, reduces transaction costs, and solves the problem of information asymmetry in traditional financing models for supply chain finance participants, providing a platform with high information transparency. By utilizing blockchain technology, the validity of transactions can be effectively confirmed, leading to decreased financing expenses for small and medium-sized enterprises (SMEs), lowered assessment charges for key companies, and improved oversight efficiency for financial institutions. The integration of blockchain technology not only enhances the efficiency of operations across the entire supply chain finance system but also positively promotes its sustainable development. This article explores the persistent equilibrium condition in the market for supply chain finance financing operations and analyzes it using evolutionary game theory, aiming to verify whether the integration of blockchain technology can help reduce the entry costs for various game participants in supply chain finance. Meanwhile, it meticulously analyzes the behaviors of each game participant, aiming to provide strategic recommendations for relevant departments to advance the progression of supply chain financing further.

## 2.related research

### 2.1 Financial Supply Chain Management

In the study of supply chain financing, the relationships between enterprises and various participants occupy a central position. According to existing research findings, the essence of supply chain financing lies in deeply integrating capital flows and logistics through close collaboration, thereby enhancing the operational efficiency of the entire supply chain. During this process, tacit cooperation and interaction among various entities are crucial. Timme et al. defined the notion of supply chain financing with a focus on the collaborative efforts between supply chain members and external financial service providers. This collaboration is based on the common pursuit of supply chain goals, integrating resources and advantages from all parties to create greater supply chain value as a whole. Chen and Hu described the functions of supply chain finance from the perspectives of banks and capital-constrained enterprises, highlighting its role as a financial innovation that acts as a conduit between banks and enterprises facing capital shortages. Through the utilization of supply chain financing, banks can acquire a deeper understanding of enterprises' operating statuses and risk profiles, enabling them to offer more customized financial services. Huang Shaoqing analyzed the participants and their roles in supply chain finance from a dynamic perspective, asserting that the participants in supply chain financing encompass financial institutions, third-party logistics firms, key enterprises, etc. These entities each play their unique roles in driving the evolution of supply chain finance.

**2.2 Blockchain Technology**

In 2008, Yap K Y et al. were the first to propose the concept of blockchain technology, identifying it as the cornerstone of distributed digital currencies such as Bitcoin. Essentially, blockchain technology is a distributed database that ensures smooth information transmission between data blocks. This innovation initiated in-depth research on blockchain technology by scholars both domestically and internationally. Queiroz M M constructed a corresponding model based on practical cases of blockchain technology in the logistics sectors of India and the United States, and used the PLS-SEM method for model evaluation, confirming the effectiveness of blockchain technology in the logistics field. Shi Aiwu and Han Chao et al. introduced blockchain technology into the telemedicine field, designing innovative technologies such as the SMSSS-CA signature mechanism and NDN naming data network-integrated smart contracts, and found that the SMSSS-CA signature mechanism can significantly reduce the operational and maintenance costs of existing hospitals. Zhang Lei et al. proposed a new framework for mutual trust and co-governance in property management based on blockchain technology, aiming to address issues such as insufficient transparency, lack of professional competence, absence of consensus mechanisms, and periodic crises in property management, ensuring its healthy operation. He Chengying et al. emphasized the crucial role of cutting-edge technologies such as blockchain and artificial intelligence in the development of the metaverse, believing that technologies like virtual reality and blockchain can leverage their technical advantages to promote industrial transformation and upgrading while maintaining the characteristics of existing industries.

**2.3 Evolutionary Game Theory**

The theory of evolutionary games, an important branch of game theory, traces its origins to Maynard Smith's pioneering work in 1973. Later on, in 1974, Smith and Price further developed the theoretical foundation, introducing the fundamental notion of evolutionarily stable strategy (ESS). In 1978, Taylor and Jonker introduced the notion of replicator dynamics, laying the foundation for constructing the basic dynamic models of evolutionary games and accelerating the progression or advancement of the theory. Evolutionary game theory focuses on the dynamic evolutionary process driven by imitation, learning, and environmental adaptation among individuals within a population. In evolutionary game applications, two core concepts—the replicator dynamic equation and the evolutionary stable strategy—occupy central positions. Taylor and Jonker were the first to introduce the replicator dynamic model in 1978, which was later expanded by Cressman, pointing out that the replicator dynamic equation can ensure the consistency between evolutionary stable strategies and evolutionary equilibria, accurately depicting the dynamic changes in individual strategies within the system. Therefore, the replicator dynamic equation has become a widely used dynamic selection model in evolutionary game practice, often expressed mathematically in the form of differential equations. On the other hand, Smith and Price initially defined the evolutionary stable strategy in 1974, which was further enriched by Hirshleifer through the introduction of the concept of evolutionary equilibrium, achieving an organic integration of dynamic processes and static concepts. Friedman's assertion that "evolutionary equilibria must be Nash equilibria" injected new vitality into evolutionary equilibrium research. Friedman's elaboration on multi-population evolutionary stable strategies has gained wide acceptance in the academic community, and the research in this paper is built upon this solid theoretical foundation.

**2.4 Financing Methods**

The financing methods within supply chain finance can be categorized into two types: external financing and internal financing. External financing constitutes a key source of funds for capital-needy enterprises, encompassing five main forms: bank loans, credit guarantees, inventory pledges, factoring financing, and order financing. a. Bank Loans: In the supply chain finance system, banks provide corresponding-scale loan support to enterprises based on their credit status and risk assessments. b. Credit Guarantees: Another key external financing channel in supply chain finance, where financially abundant enterprises typically provide credit guarantees to capital-constrained enterprises, promising to bear repayment responsibilities if the latter fails to repay on time, thereby helping them obtain the required funds. c. Inventory Pledges: Inventory pledges allow capital-constrained enterprises to use their inventory as collateral to obtain loans from financing institutions. d. Factoring Financing: A financing method where suppliers transfer their accounts receivable to third-party financing institutions to obtain immediate cash flow. e. Order Financing: An important financing mode in supply chain finance that allows buyers to obtain loans by selling their purchase orders to third-party institutions.

## 3. Method

### 3.1.Construction of the Supply Chain Finance Operating Framework and Analysis of Relationships

The uniqueness of blockchain technology lies in its "perpetual traceability and immutability," which, when applied in the field of supply chain finance, can significantly enhance the authenticity and reliability of data information. In the platform architecture of supply chain finance, all participating entities are required to submit necessary documents and complete the membership registration process. SMEs (small and medium-sized enterprises), core enterprises, and financial institutions, as the core participants in supply chain finance, have their important information recorded in detail and accurately within the blockchain system. The transparency of blockchain technology and its ability to ensure the authenticity of information on the chain grant every participant on the platform the right to access the real information of other participants, thereby ensuring open sharing and a high degree of transparency of information. As shown in Figure 1 below, when facing financing needs, SMEs can directly initiate financing applications based on the credit information transmitted by core enterprises on the supply chain finance platform. Compared to the traditional physical asset pledge financing model in supply chain finance, this financing method has shifted to digital financing on the chain, which not only significantly enhances security but also improves efficiency. This transformation helps reduce the financing thresholds for SMEs, thereby effectively addressing the common issue of financing difficulties for SMEs and promoting their healthy development. For core enterprises, due to the transparency and immutability of information, they can directly obtain and assess the qualifications and credit status of SMEs through the platform without investing significant human resources to review the credit history of SMEs; they only need to read relevant data on the platform, thereby simplifying business processes and reducing assessment costs. For financial institutions, they can utilize the information on the platform to assess the credit situation of core enterprises, verify the authenticity of the transaction data provided by them, and make lending decisions based on the assessment results. At the same time, financial institutions can also monitor the business dynamics of SMEs and core enterprises in real-time through the platform during the loan process, thereby reducing assessment and supervision costs. In summary, from a theoretical perspective, thanks to the information transparency of blockchain technology, blockchain-driven supply chain finance can more effectively reduce the cost expenditures of all participating parties compared to traditional models.

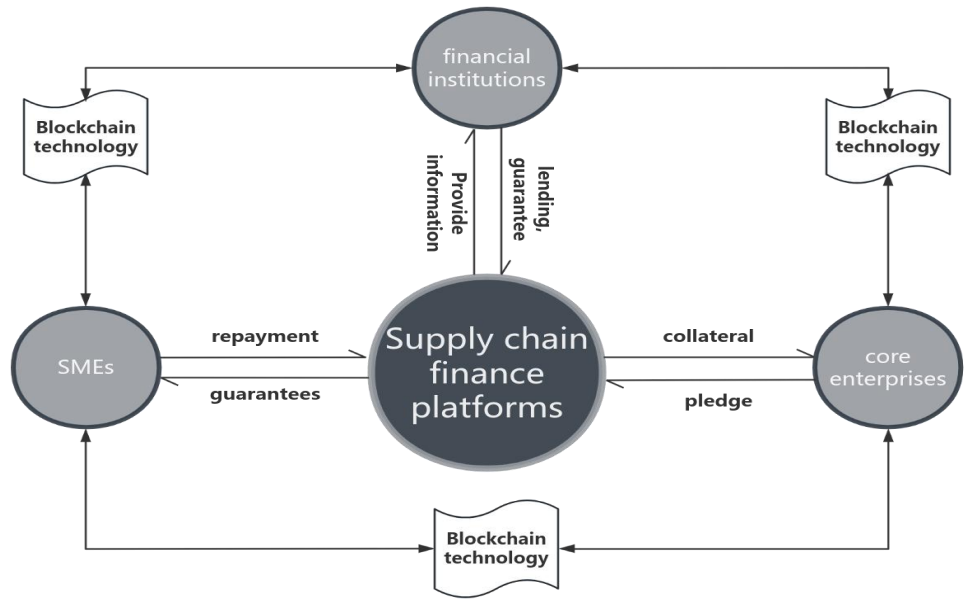

Figure 1 Supply chain finance operation framework enabled by blockchain

In the supply chain finance system, core enterprises, small and medium-sized enterprises (SMEs), and financial institutions constitute the three major participants. SMEs, as the demand side for funds, bear financing costs such as physical collateral during the financing process. If financing is successful, they must follow the loan agreement and repay the principal and interest on time. Core enterprises occupy a central position in this system and serve as credit endorsers, having a significant impact on SMEs' financing applications. In response to SMEs' financing requests, core enterprises rigorously review their qualifications and credit situations. Specifically, SMEs face two options: "accept financing" or "forgo financing." If they choose to "accept financing," they need to apply for loans using means such as physical collateral to bolster operating funds and meet business needs. Under this path, SMEs also need to bear search costs, collateral costs, and interest expenses. If they choose to "forgo financing," business processes remain unchanged. Core enterprises face the choice of "providing a guarantee" or "refusing a guarantee." If they choose to "provide a guarantee," they must thoroughly assess SMEs' credit situations, including their operating status and past credit records, which will incur assessment costs and expose them to potential bad debt risks, but they can also earn interest returns through the guarantee. If they choose to "refuse a guarantee," the core enterprise's earnings remain unchanged. Financial institutions also face the decision of "cooperating" or "not cooperating." Once they decide to "cooperate," financial institutions need to bear the assessment costs for core enterprises and supervision costs during the financing process, but they will also obtain additional earnings, primarily from the interest-bearing repayments of core enterprises. If they choose "not to cooperate," although they maintain their original earnings, they miss out on the opportunity to establish partnerships with core enterprises and other members of the supply chain.

### 3.2 Game Model Construction

Before constructing the model, we need to preset the following: The integration of blockchain technology into the supply chain finance system can reduce the financing costs across the entire chain. Specifically, the reduction in financing costs for SME entity α during fund-raising with the assistance of blockchain technology is denoted as $m_1$, the reduction in assessment expenses for core enterprise entity β is denoted as $m_2$, and the decrease in assessment costs for core enterprise β and monitoring costs of transaction processes for financial institution entity γ is denoted as $m_3$. Here, we adopt the method of replicator dynamics equations to conduct game analysis. Based on the payment and revenue matrix shown in previous Table 1, which includes the three major participants, we set $E_x$ to represent the expected revenue of SME entity α when choosing the "accept financing" strategy, and $E_{1-x}$ to represent its expected revenue when choosing the "do not finance" strategy. $E_s$ represents the average expected revenue of SME entity α under both strategies. Similarly, $E_y$ represents the expected revenue of core enterprise entity β when choosing the "provide guarantee" strategy, and $E_{1-y}$ represents its expected revenue when choosing the "do not provide guarantee" strategy. $E_c$ represents the average expected revenue of core enterprise entity β under both strategies. Based on this, the following formulas are derived. Additionally, we assume that the expected revenue of financial institution entity γ when "agreeing to cooperate" is $E_z$, and the expected revenue when "refusing to cooperate" is $E_{1-z}$. $E_d$ represents the average expected revenue of financial institution entity γ under both strategies, and based on this, the following formulas are derived:

$$E_x = R_1 - C_1 + m_1 + yz[r - \theta(K + I_1)] \quad \text{(Equation 1)}$$

$$E_{1-x} = yzR_1 + y(1-z)R_1 + z(1-y)R_1 + (1-z)(1-y)R_1 = R_1 \quad \text{(Equation 2)}$$

$$E_s = xE_x + (1-x)E_{1-x} = R_1 - x(C_1 - m_1) + xyz[r - \theta(K + I_1)] \quad \text{(Equation 3)}$$

$$F(x) = \frac{dx}{dt} = x[E_x - E_s] = x(1-x)[yzr - yz\theta(K + I_1) - C_1 + m_1] \quad \text{(Equation 4)}$$

Similarly, the following formulas can be derived:

$$E_c = yE_y + (1-y)E_{1-y} = R_2 - y(C_2 - m_2) + xyz[I_1 - I_2 + S - (1-\theta)(K + I_1)]$$

(Equation 5)

$$G(y) = \frac{dy}{dt} = y[E_y - E_c] = y(1-y)[xz(I_1 - I_2 + S) - xz(1-\theta)(K + I_1) - C_2 + m_2]$$

(Equation 6)

$$E_d = zE_z + (1-z)E_{1-Z} = R_3 - z(C_3 - m_3) + xyzI_2 \qquad \text{(Equation 7)}$$

$$H(z) = \frac{dz}{dt} = z[E_z - E_d] = z(1-z)(xyI_2 - C_3 + m_3) \qquad \text{(Equation 8)}$$

| | | core enterprises β | Cooperation z | refusal to cooperate 1-z |
|---|---|---|---|---|
| SMEs α | Financing x | guarantees y | $R_1+r-\theta(K+I_1)-C_1+m_1$<br>$R_2+I_1-C_2+m_2-(1-\theta)(K+I_1-S)-I_2$<br>$R_3+I_2-C_3+m_3$ | $R_1-C_1+m_1$<br>$R_2-C_2+m_2, R_3$ |
| | | refusal to guarantee 1-y | $R_1-C_1+m_1$<br>$R_2$<br>$R_3-C_3+m_3$ | $R_1-C_1+m_1$<br>$R_2$<br>$R_3$ |
| | refusal to finance 1-x | Guarantees y | $R_1, R_2-C_2+m_2$<br>$R_3-C_3+m_3$ | $R_1, R_2-C_2+m_2$<br>$R_3$ |
| | | refusal to guarantee 1-y | $R_1, R_2$<br>$R_3-C_3+m_3$ | $R_1, R_2, R_3$ |

Table 1 Payoff Matrix

To solve the replicator dynamics equations derived from the model mentioned above, we set each of the three expressions within the system of equations to zero, thereby obtaining nine locally equilibrium states contained within the model. To explore the local stability characteristics of these equilibrium states, we construct the Jacobian matrix of the replicator dynamics equations:

$$J = \begin{bmatrix} F_{11} & F_{12} & F_{13} \\ F_{21} & F_{22} & F_{23} \\ F_{31} & F_{32} & F_{33} \end{bmatrix} = \begin{bmatrix} \frac{\partial F(x)}{\partial x} & \frac{\partial F(x)}{\partial y} & \frac{\partial F(x)}{\partial z} \\ \frac{\partial G(y)}{\partial x} & \frac{\partial G(y)}{\partial y} & \frac{\partial G(y)}{\partial z} \\ \frac{\partial H(z)}{\partial x} & \frac{\partial H(z)}{\partial y} & \frac{\partial H(Z)}{\partial z} \end{bmatrix} \qquad \text{(Equation 9)}$$

Based on relevant information, Figure 2 depicts the progression of the supply chain finance financing game model bolstered by blockchain technology, in which X, Y, and Z signify respectively the likelihood of small and medium-sized enterprises (SMEs) α adopting the "agree to finance" strategy, the likelihood of core enterprises β opting for the "offer guarantee" strategy, and the likelihood of financial institutions γ choosing the "engage in cooperation" strategy. Thanks to blockchain technology, the supply chain finance financing game model showcases two Evolutionary Stable Strategies (ESS) conditions, and the final evolutionary outcome of this model is contingent upon the initial strategic choice probabilities of the three participating entities. In its starting phase, when the probabilities of game participants embracing proactive financing strategies are elevated, indicating fewer occurrences of SMEs "declining finance," core enterprises "declining guarantees," and financial institutions "declining collaboration," the model is inclined to converge towards the ESS state of (1,1,1). Alternatively, if the probabilities of SMEs "declining finance," core enterprises "declining guarantees," and financial institutions "declining collaboration" are elevated, the likelihood of the model converging to the alternate ESS state will increase accordingly.

| Loan term | benchmark interest rate |
|---|---|
| within one year | 4.35% |
| one to five years | 4.75% |
| above five years | 4.9% |

Table2 Loan Benchmark Interest Rate

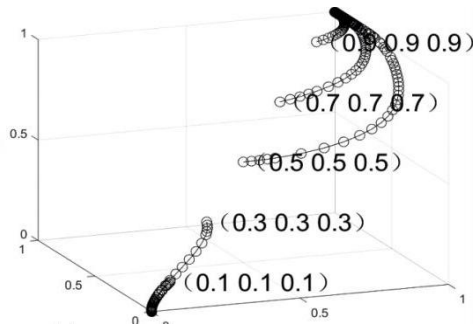

Figure 2 Evolutionary Path of Game Models

## 4.Results and discussion

Based on the assessment, both prior to and following the incorporation of blockchain technology, the supply chain finance game model demonstrates identical localized equilibrium points, specifically including: $E_1(0,0,0)$, $E_2(1,0,0)$, $E_3(0,1,0)$, $E_4(0,0,1)$, $E_5(0,1,1)$, $E_6(1,0,1)$, $E_7(1,1,0)$, and $E_8(1,1,1)$. However, when assessing the stability of these equilibrium points, the conditions to be considered differ, especially for the equilibrium point $E_8(1,1,1)$. The conditions for it to become an ESS (Evolutionarily

Stable Strategy) in the two financing game models are vastly different. Specifically, in the original model, the conditions for E8 to become an ESS are: $r > 1 + \theta(I_1 + K)$, $S + \theta(I_1 + K) > I_2 + K + C_2$, and $I_2 > C_3$. In the model integrated with blockchain technology, the conditions for it to become an ESS depend on different combinations of parameters. Consequently, the implementation of blockchain technology reduces the barriers for the supply chain finance game model to attain Pareto optimality, enabling smoother execution of tripartite transactions.

## 5.Conclusion

During the evolution of supply chain finance, issues such as information asymmetry and inefficient credit transmission have always served as core obstacles hindering its smooth and sound development. Utilizing the theoretical foundations of supply chain finance and evolutionary game theory, this research establishes a novel supply chain finance financing game model that incorporates blockchain technology. By establishing and analyzing replicator dynamics equations, we thoroughly discuss the strategic choices and behavioral characteristics of game participants in this context. Furthermore, we utilize the Jacobian matrix to analyze the local stability of the obtained equilibrium points, and accordingly derive evolutionarily stable strategies under various scenarios. The supply chain finance financing game model integrated with blockchain technology demonstrates characteristics of Pareto improvement, opening up broader prospects for the prospective advancement of supply chain finance.